\documentclass[preprint,prb,aps,amsmath,amssymb,color,groupedaddress]{revtex4}
\usepackage{stmaryrd}
\usepackage{amsmath}
\usepackage{amssymb}
\usepackage{graphicx}
\usepackage{dcolumn}
\usepackage{bm}
\begin{document}	
\begin{titlepage}
\title{Vector Potential and Surface Magnetic Field in Magnetoelectric Antiferromagnetic Materials}
\author{Zeyu Jiang}
\author{Damien West}
\author{Shengbai Zhang}
\email{zhangs9@rpi.edu}
\affiliation{Department of Physics, Applied Physics and Astronomy, Rensselaer Polytechnic Institute, Troy, NY, 12180, USA}
\date{\today}
\begin{abstract}
A general formula for the average vector potential of bulk periodic systems is proposed and shown to set the boundary conditions at magnetic interfaces. For antiferromagnetic materials, the study reveals a unique relation between the macroscopic potential and the orientation-dependent magnetic quadrupole, as a result of the different crystalline and magnetic symmetries. In particular, at surfaces and interfaces of a truncated bulk without inversion and time-reversal symmetries, the average vector potential exhibits a discontinuity, which results in an interfacial magnetic field. In general, however, due to the surface and interface electronic and atomic relaxations, additional magnetization may result. For the experimentally-observed magnetoelectric antiferromagnets, in particular, our symmetry analysis suggest that the relaxation effects could well be a system response to the presence of such a potential discontinuity. 
\end{abstract}
\maketitle
\draft
\vspace{2mm}
\end{titlepage}

Antiferromagnetism with fully compensated spin sublattices is an active research field with growing interest for their potential applications in spintronics, magnetoelectrics, magnetic control, and high performance storage\cite{Baltz}. Typically, a bulk antiferromagnetic system possesses zero total magnetization, thus presenting no macroscopic magnetic field and, as such, it is insensitive to an external field. The disappearance of the macroscopic magnetization protects the information stored in an antiferromagnetic material from ambient perturbations and guarantees an extremely-high element density\cite{Jungwirth}. However, the compensated spin arrangement also makes any direct probing and manipulating of the magnetic configuration difficult, hence hindering the use of antiferromagnetism in primary device applications. 

To overcome this drawback, recently considerable efforts have been made to the study of magnetoelectric antiferromagnets\cite{Eerenstein,Kusob,Oh,Mong}, which exhibit an electric control of the magnetization, and as such they could serve as a promising alternative to traditional multiferroics. Generally speaking, the magnetoelectric effect refers to the ability of controlling the magnetization (or electric polarization) through an electric (or magnetic) field, where a violation of both spatial inversion and time-reversal symmetries leads to a non-vanishing coupling between the magnetization and electric polarization\cite{Zarzuela}. The electric polarization can in turn be controlled and tuned by a variety of technologies. Note that the combined inversion and time-reversal symmetry can still be preserved in such systems, which ensures the absence of a net magnetization or polarization in the equilibrium bulk state\cite{Gaol}. However, the presence of a surface can lead to a finite, uncompensated magnetization distributed on the surface, even in a perfect magnetoelectric antiferromagnet\cite{Belashchenko}. This surface magnetization is especially important in a number of applications such as in multiferroic devices, in magnetic proximity effect, and in exchange bias. 

For the well-known magnetoelectric material Cr$_2$O$_3$, a macroscopic magnetic field outside the sample has been observed\cite{Astrov}, which has a distance dependence of 1/R$^4$. Such a field is explained in terms of a surface magnetization analogous to the macroscopic electric field from a surface polarization\cite{Andreev,Wu}. However, there are also suggestions attributing the 1/R$^4$ field to bulk magnetic quadrupole from which a 1/R$^4$ law can be deduced\cite{Astrov,Dzyaloshinskii}. In a recent work\cite{Choe2018,Choe2019}, we showed that the bulk electric quadrupole will trigger a discontinuity in the average electrostatic potential, which leads to the formation of a surface macroscopic electric field even in materials without any macroscopic electric dipole. This surface electric field cannot be described by any surface properties, but is solely determined by the bulk electric quadrupole. For the magnetic quadrupole, the physical significance can be greater than that of the electric quadrupole, as thermodynamic analysis has shown that the derivative of the magnetic quadrupole with respect to the chemical potential is the origin for the aforementioned electromagnetic effect\cite{Gaos,Shitade}. The anti-symmetric part of the magnetic quadrupole is directly related to a magnetic toroidal moment\cite{Dubovik,Spaldin,Kopaev}, a uniform arrangement of which forms a fourth ferroic order, dubbed as ferotoroidicity\cite{Eerenstein,Fiebig,Gnewuch}. These raise the question: can bulk magnetic quadrupole also play a role in the manifestation of the experimentally-observed magnetism? 

In this paper, we derive from the Biot-Savart Law a general expression of the (microscopic) average vector potential for antiferromagnetic systems and find it to be solely determined by the bulk magnetic quadrupole. As such, it is found that the average vector potential encounters a discontinuity at the interface/surface of dissimilar materials, leading to a (macroscopic) magnetic field as a manifestation of this pure intrinsic-bulk effect. Using two simplified models as propotypical antiferromagnetic systems, we illustrate the aforementioned interfacial magnitization. Our results may reconcile the controversy on the origin of the observed magnetic field in proximty to magnetoelectric antiferromagnet surfaces. Furthermore, consideration of spatial inversion and time-reversal symmetry supports the notion that this intrinsic effect is the root cause for the proximity field with atomic and electronic relaxation yielding only secondary contributions.

In electrodynamics, the scalar potential $\varphi$ and vector potential $\textbf{A}$ are written as
\begin{equation}
\label{eqn:phi'}
\varphi' = \varphi - \frac{\partial \Lambda}{\partial t}
\end{equation}
\begin{equation}
\label{eqn:a'}
\textbf{A}' = \textbf{A} + \nabla \Lambda,
\end{equation}
where $\Lambda(\textbf{r}, t)$ is an arbitrary function of time and position, associated with gauge transformation. If we concentrate on (stationary) ground-state properties, all the quantities in Eq. \ref{eqn:phi'} and \ref{eqn:a'} are only functions of $\textbf{r}$. In particular, $\frac{\partial \Lambda}{\partial t}=0$, so the gauge arbitrariness in $\varphi$ is removed up to a constant $\phi_{0}$ (common energy reference). For a finite system, vacuum level $\phi_{0}=0$ is often used as the reference. For an infinite system, however, the identification of $\phi_{0}$ becomes an issue, despite its crucial importance in the physics of interface properties, such as charge transfer, Schottky barrier height, and band alignment. In recent work\cite{Choe2018}, we showed that the maximum of the planar-averaged electrostatic potential equals to $\phi_{0}$ and may be used to align the band structures of bulk systems (or alternately, truncated bulk with ideal surfaces). While actual heterostructures have the additionl effect of electronic and atomic relaxation, this allowed for the interfacial properties to be divided into two parts: one of which is determined purely by bulk properties with the other involving interfacial relaxations due to system energy minimization. 

The situation can be more complicated for vector potential $\textbf{A} $, because $\nabla \Lambda$ holds a large degree of freedom even when $\frac{\partial \Lambda}{\partial t}=0$. In spite of this arbitrariness, the potential difference across a magnetic interface can be physically defined. To see this, we draw an analogy between the electric potential $\varphi$ and the vector potential $\textbf{A} $. Typically, $\varphi$ is given in a Coulomb gauge, which differs from $\varphi'$ by $\phi_{0}$. Similarly, $\textbf{A}$ in the Coulomb gauge differs from $\textbf{A}'$ by $\nabla \Lambda$. Mathematically, this means 
\begin{equation}
\label{eqn:phi'_integrated}
\varphi' = \frac{1}{4\pi\epsilon_{0}} \int_{\infty} d^{3}{r'} \frac{\rho(\textbf{r}')}{|\textbf{r}-\textbf{r}'|} + \phi_{0} 
\end{equation}
and
\begin{equation}
\label{eqn:A'_integrated}
\textbf{A}' = \frac{\mu_{0}}{4\pi} \int_{\infty} d^{3}{r'} \frac{\textbf{J}(\textbf{r}')}{|\textbf{r}-\textbf{r}'|} + \nabla \Lambda(\textbf{r}).
\end{equation}
By comparing Eqs. \ref{eqn:phi'_integrated} and \ref{eqn:A'_integrated}, it is clear that in the magnetostatic limit $\nabla \Lambda$ assumes the role of magnetic vacuum level, similar to the role of $\phi_{0}$ in the electrostatic limit. In other words, no matter how complex $\nabla \Lambda$ are, they should ``align'' with each other at the magnetic interface, thereby eliminating most of the arbitrariness associated with $\nabla \Lambda$ and contributing nothing to the potential difference. 

We begin with the derivation of $\overline{\textbf{A}}$ in bulk materials. In the case of $\varphi$, charge density $\rho(\textbf{r}')$ is well-defined, which shares the same translational symmetry as the Bravais lattice. To obtain $\overline{\varphi}$, one simply takes the average of Eq. \ref{eqn:phi'_integrated} in one unit cell. However, for a bulk system, the choice of the unit cell is also perceived arbitrary, which was fortunately resolved in Ref.\cite{Choe2018}, based on the physical requirement that electric dipole for the truncated bulk should vanish. In contrast, a periodic current density $\textbf{J}(\textbf{r}')$ is not as straightforward. To proceed, we refer $\textbf{J}(\textbf{r}')$ as the microscopic circulating currents coming from a periodic array of magnetic (dipole) moments, which may originate from either a spin or an orbital angular motion. This way, $\textbf{J}(\textbf{r}')$ also preserves the translational symmetry of the Bravais lattice and can thus be averaged over one unit cell. Note that, to simultaneously satisfy the zero electric dipole requirement, the same unit cell in Eq. \ref{eqn:phi'_integrated} should be used in Eq. \ref{eqn:A'_integrated}. Also, the use of $\textbf{J}(\textbf{r}')$ here is purely an intermediate step. In the end, we will seek only physical observables that have real physical meaning, such as the magnetic field generated by the magnetic moments. 

The value of $\overline{\textbf{A}}$ is obtained by averaging over a unit cell
\begin{equation}
\label{eqn:Abar_int_unit}
\overline{\textbf{A}} = \frac{\mu_{0}}{4\pi\Omega} \int_{\Omega}d^{3}{r} \int_{\infty}d^{3}{r'} \frac{\textbf{J}(\textbf{r}')}{|\textbf{r}-\textbf{r}'|},
\end{equation}
where $\Omega$ is the volume of the unit cell. Because $\textbf{J}(\textbf{r}')$ in Eq. \ref{eqn:Abar_int_unit} is a periodic function (i.e., $\textbf{J}(\textbf{r}')= \textbf{J}(\textbf{r}'+\textbf{R}_{I})$, the second integration over $r’$ over an infinite space may be rewritten as an integration over just one unit cell, summed over the cell index $-\infty\le I \le\infty$, namely,
\begin{equation}
\label{eqn:Abar_sum}
\overline{\textbf{A}} = \frac{\mu_{0}}{4\pi\Omega} \sum_{I} \int_{\Omega}d^{3}{r} \int_{\Omega}d^{3}{r'} \frac{\textbf{J}(\textbf{r}')}{|\textbf{r}-\textbf{r}' - R_{I}|},
\end{equation}
where $R_{I}$ is the position of the $I$th unit cell. Next, if we associate sum over $I$ with the first integration over $\textbf{r}$ also within one unit cell, Eq. \ref{eqn:Abar_sum} may be rewritten as
\begin{equation}
\label{eqn:A_unfold}
\begin{aligned}
\overline{\textbf{A}} &= \frac{\mu_{0}}{4\pi\Omega} \int_{\infty}d^{3}{r} \int_{\Omega}d^{3}{r'} \frac{\textbf{J}(\textbf{r}')}{|\textbf{r}-\textbf{r}'|} \\
&=\frac{\mu_{0}}{4\pi\Omega} \int_{\infty}d^{3}{r} \textbf{A}_{\Omega}(\textbf{r}).
\end{aligned}
\end{equation}
Here, $\textbf{A}_{\Omega}(\textbf{r})$ is the vector potential generated by the current density in a single unit cell. The scheme in arriving at Eq. \ref{eqn:A_unfold} from Eq. \ref{eqn:Abar_int_unit} is exactly the same as the potential-unfolding scheme developed in Ref.\onlinecite{Choe2019} for the average electric potential. This can be easily seen if we express $\overline{\textbf{A}}$ as three independent integrals in respective orthogonal orientations. 

Next, we define the unit cell current density
\begin{equation}
\label{eqn:J0_def}
\textbf{J}_{0}(\textbf{r}')=\left\{
\begin{aligned}
\textbf{J}(\textbf{r}'),& & \textbf{r}' \in \Omega \\
\textbf{0},& & \textbf{r}' \notin \Omega
\end{aligned}
\right.,
\end{equation}
which is mathematically easier to handle than the extended $\textbf{J}$ over the entire space. In doing so, the integration over $\textbf{r}'$ in the middle expression in Eq. \ref{eqn:A_unfold} can also be extended to infinity,
\begin{equation}
\label{eqn:Abar_J0}
\overline{\textbf{A}} = \frac{\mu_{0}}{4\pi\Omega} \int_{\infty}d^{3}{r} \int_{\infty}d^{3}{r'} \frac{\textbf{J}_{0}(\textbf{r}')}{|\textbf{r}-\textbf{r}'|}.
\end{equation}

\begin{figure}[tbp]
\includegraphics[width=0.95\columnwidth]{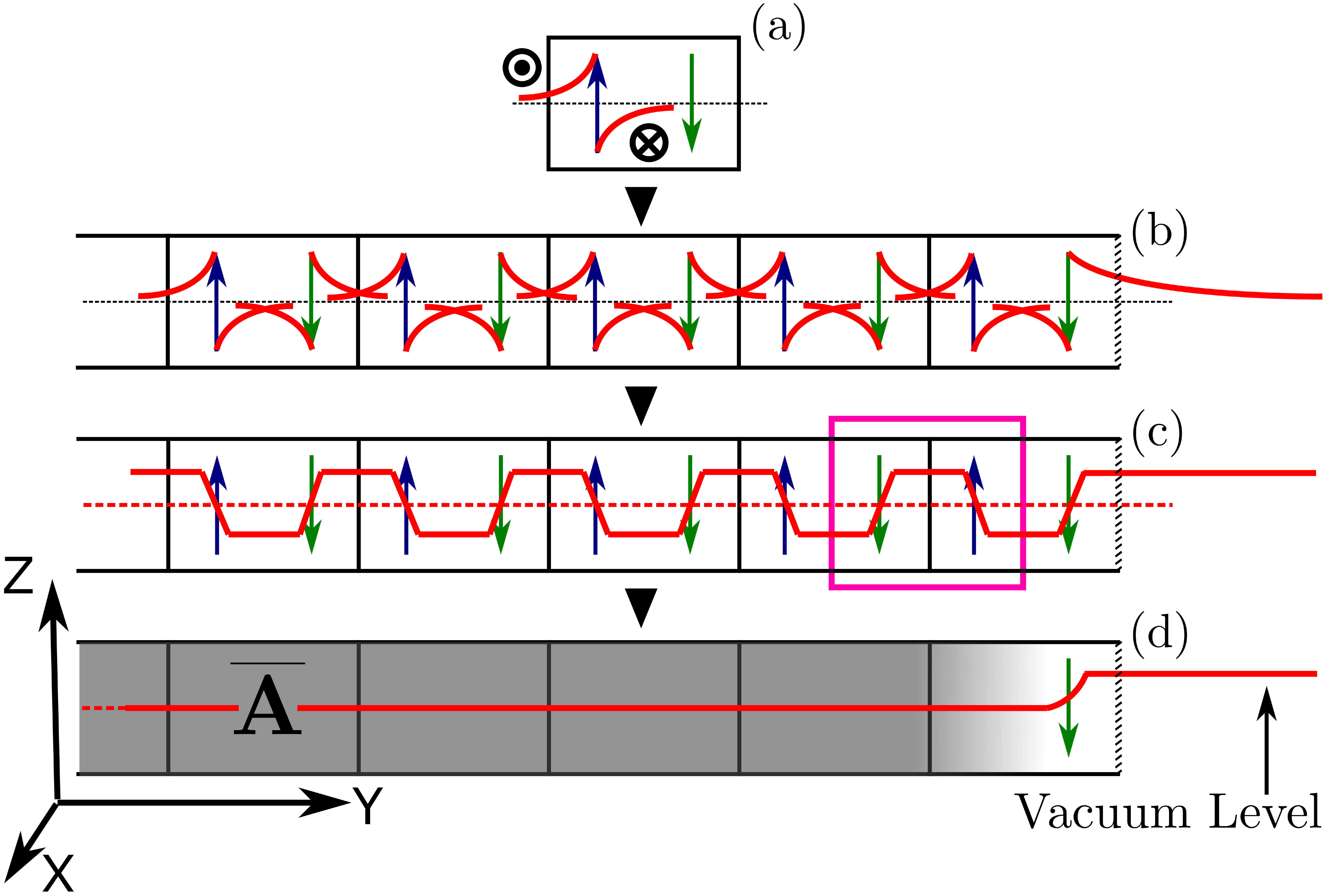}
\caption{\label{fig:fig1} (Color online) Schematic illustrations of $\overline{\textbf{A}}$ in proximty to the surface (right). (a) The potential produced by a single magnetic moment, the blue up-arrow. Note that the resulting vector potential is in the $\textbf{e}_x$ direction and the red curve indicates the amplitude with positive being out of the page $\odot$ and negative being into the page $\otimes$. There are two magnetic moments of opposite directions (i.e., the blue and green arrows) in each unit cell of the antiferromagnetic system. (b) A superposition of vector potentials of individual magnetic moments. (c) XZ-plane-averaged vector potential as a function of Y. Each unit cell contains a net magnetic quadrupole in the X-direction, which contributes only a constant potential to regions outside the unit cell (indicated by the purple frame). (d) The macroscopically averaged vector potential. Inside the bulk, it is a constant but raises near a surface due to the unpaired outmost (green) magnetic moment. On the opposite surface, on the other hand, one can expect an unpaired (blue) magnetic moment, causing a similar effect.}
\end{figure}

Note that the result of Eq. \ref{eqn:Abar_J0} will generally depend on the order of the integration\cite{Choe2019}. In the reciprocal space, this peculiar behavior manifests itself as along which direction to take the $\textbf{q}\rightarrow\textbf{0}$ limit. Because $\textbf{J}_{0}(\textbf{r}')$ is non-periodic, a Fourier transform will result in a continuous $\textbf{q}$, as opposed to having a sum over discrete \{$\textbf{G}$\} vectors. We need $\textbf{A}(\textbf{q})$ to be a continuous function here, as strictly speaking $\textbf{A}(\textbf{q=0})$ is an ill-defined quantity. To carry out the integration, we follow the standard procedure: first we Fourier transform the Coulomb interaction $1/|\textbf{r}-\textbf{r}'|$ to obtain
\begin{equation}
\label{eqn:Abar_J0_q}
\overline{\textbf{A}} = \frac{\mu_{0}}{4\pi\Omega} \int_{\infty}d^{3}{r} \int_{\infty}d^{3}{r'} \int_{\infty}d^{3}{q} \frac{4\pi}{(2\pi)^3} \frac{\textbf{J}_{0}(\textbf{r}')}{|\textbf{q}|^2} e^{i\textbf{q}\cdot(\textbf{r}-\textbf{r}')} .
\end{equation}
Next, we perform the integrations in Eq. \ref{eqn:Abar_J0_q} in the following order: $\textbf{r}'$, $\textbf{r}$, and then $\textbf{q}$, which yields
\begin{equation}
\label{eqn:Abar_J0_q_short}
\overline{\textbf{A}} = \frac{\mu_{0}}{\Omega} \lim_{\textbf{q}\to\textbf{0}} \frac{\textbf{J}_{0}(\textbf{q})}{|\textbf{q}|^2}.
\end{equation}
The physics of the directional dependence in Eq. \ref{eqn:Abar_J0_q_short} originates from the surface anisotropy even of truncated bulk\cite{Choe2018}. This has been demonstrated for $\overline{\varphi}$ in Ref. \cite{Choe2019}. It was shown that the orientation-dependence reflects the nonequivalence of the outermost-layer electric dipoles, as a result of cutting the bulk crystal at crystallographic planes (or chemical bonds) at different orientations. Similarly, the outermost-layer surface magnetization leads to a macroscopic discontinuity in the vector potential, as shown in Fig. 1. The magnitude of the discontinuity is also generally anisotropic and sensitive to the orientation between the direction of the magnetic dipole and the normal direction of the surface. Note that each magnetic unit cell, denoted by a purple rectangular frame, contributes only a constant vector potential, which aligns with each other in Fig. 1(c). The net result is that the macroscopic $\overline{\textbf{A}}$ is uplifted from that of bulk by an amount set by the last, unpaired magnetic moment, as shown in Fig. 1(d). This is in complete analogy to what happens in average electric potential $\overline{\varphi}$ where it is the last unpaired electric dipole that sets the potential offset across an ideal interface.

\begin{figure}[tbp]
\includegraphics[width=0.95\columnwidth]{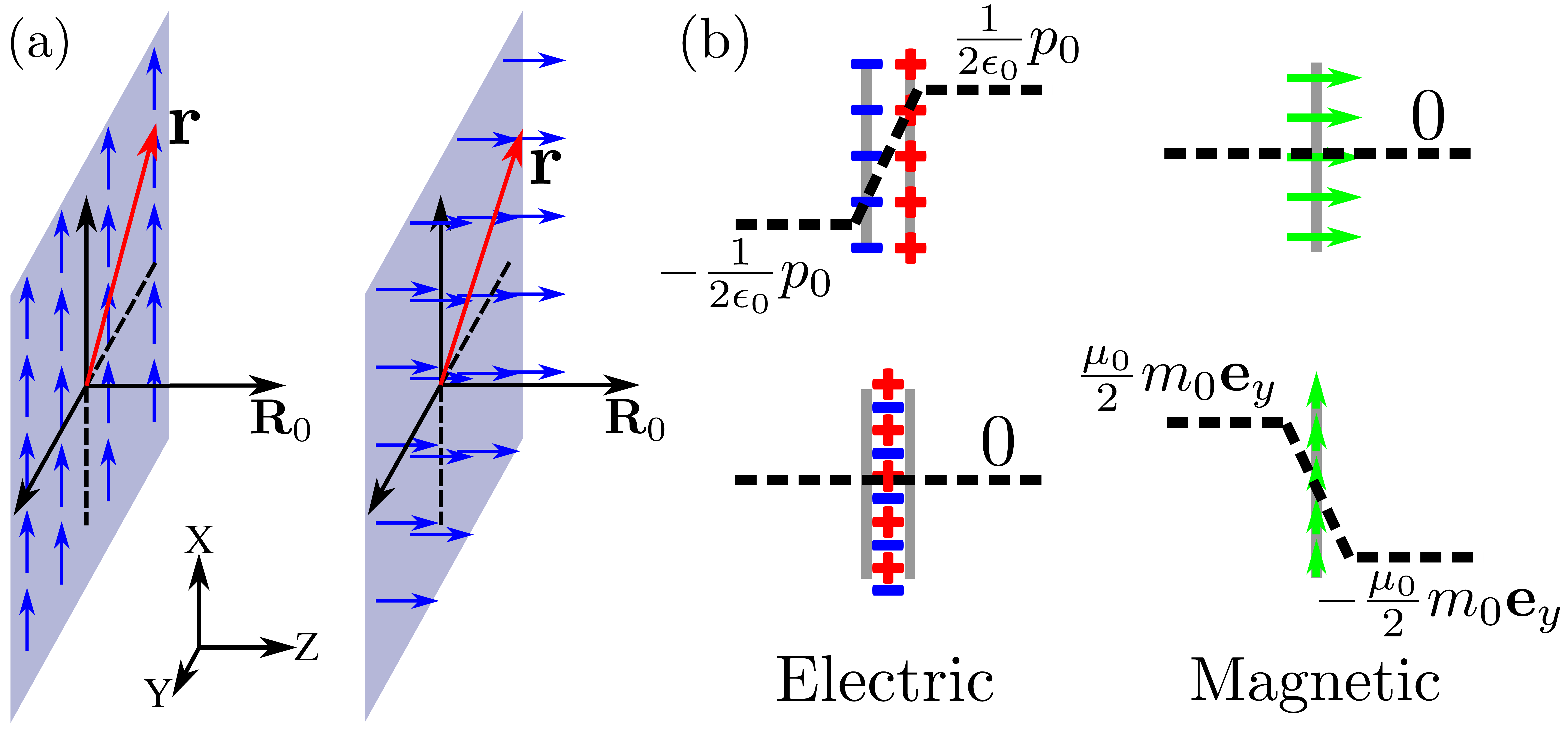}
\caption{\label{fig:fig2} (Color online) (a) A schematic illustration of a homogeneous electric dipole (or magnetic moment) distribution in an infinitely-large plane. The orientation is either parallel (left) or perpendicular (right) to the plane. Vectors $\textbf{r}$ and $\textbf{R}_{0}$ refer to the source and field points, respectively. (b) Distributions of the scalar (left two panels) and vector (right two panels) potentials across the dipole planes. Top panels are for dipoles perpendicular to the plane while lower panels are for them parallel to the plane. $-$ and $+$ symbols denote the negative and positive charges making up the electric dipoles, and the green arrow denotes the magnetic moments. Black dashed line shows schematically spatial variations of their corresponding potentials.}
\end{figure}

To better illustrate the discussion above, we consider here an infinitely-large plane with a homogeneous distribution of either electric dipoles of density $\textbf{p}_{0}$ or magnetic moments of density $\textbf{m}_{0}$, as shown in Fig. 2(a), which are either parallel or perpendicular to the plane. The scalar and vector potentials at a distance $\textbf{R}_{0}$ away from the plane may be written as 
\begin{equation}
\label{eqn:pot_plane}
\varphi_{0}(\textbf{R}_{0}) = \frac{1}{4\pi\epsilon_{0}} \int\int_{-\infty}^{+\infty} dxdy \frac{\textbf{p}_{0}\cdot\left(\textbf{R}_{0}-\textbf{r}\right)}{\left|\textbf{R}_{0}-\textbf{r}\right|^{3}}
\end{equation}
and
\begin{equation}
\label{eqn:A_plane}
\textbf{A}_{0}(\textbf{R}_{0}) = \frac{\mu_{0}}{4\pi} \int\int_{-\infty}^{+\infty} dxdy \frac{\textbf{m}_{0}\times\left(\textbf{R}_{0}-\textbf{r}\right)}{\left|\textbf{R}_{0}-\textbf{r}\right|^{3}}.
\end{equation}
Thus a $\textbf{p}_{0}$ parallel to the plane only yields a zero potential outside the plane, while a $\textbf{p}_{0}$ perpendicular to the plane yields $\varphi_{0}=\pm \frac{1}{2\epsilon_{0}} p_{0}$ on the left and right hand sides of the plane, respectively. Conversely, a $\textbf{m}_{0}$ perpendicular to the plane yields a zero vector potential outside the plane, while a $\textbf{m}_{0}$ parallel to the plane yields $\textbf{A}_{0}=\pm \frac{\mu_{0}}{2} m_{0} \textbf{e}_{y}$ on the left and right hand side of the plane, respectively. Note that in all cases both $\varphi_{0}$ and $\textbf{A}_{0}$ are independent of $\textbf{R}_{0}$. Thus, a magnetic unit cell made of pairs of opposite dipoles will represent merely a constant potential outside, in line with what has been shown in Fig. 2(b). 

At first glance, it might be a bit surprising that a discontinuity appears here, as the slab in the above derivation is merely a truncation of a bulk without any electronic or atomic relaxations. In fact, what we see here for the vector potential is similar to what we saw for the electric potential in a non-magnetic truncated slab\cite{Choe2019}, namely, it is determined by intrinsic bulk property, the magnetic quadrupole $\overleftrightarrow{\textbf{M}}$. To understand this, we take a Taylor expansion of $\textbf{J}_{0}(\textbf{q})$ in Eq. \ref{eqn:Abar_J0_q_short} up to the second order
\begin{equation}
\label{eqn:Abar_taylor}
\begin{aligned}
\overline{\textbf{A}} = \frac{\mu_{0}}{\Omega} \lim_{\textbf{q}\to\textbf{0}} \frac{1}{|\textbf{q}|^2} \Big \{ &\textbf{J}_{0}(\textbf{0}) + \textbf{q}\cdot\left[\nabla\textbf{J}_{0}(\textbf{q})\right]|_{\textbf{q}=\textbf{0}} + \\ &  \frac{1}{2!}\textbf{qq}\colon\left[\nabla\nabla\textbf{J}_{0}(\textbf{q})\right]|_{\textbf{q}=\textbf{0}} + O(q^3) \Big \}, 
\end{aligned}
\end{equation}
where the multiplication of dyadic tensor $\textbf{qq}\colon\nabla\nabla$ is defined as $\left(\textbf{q}\cdot\nabla\right)^2$. The three terms on the right-hand side of Eq. \ref{eqn:Abar_taylor} represent respectively the magnetic monopole, dipole, and quadrupole. In antiferromagnetic systems, the monopoles and dipoles naturally vanish, and the remaining quadrupole term may be written as
\begin{equation}
\label{eqn:quad_term}
\begin{aligned}
\textbf{qq}\colon\left[\nabla\nabla\textbf{J}_{0}(\textbf{q})\right]|_{\textbf{q}=\textbf{0}} &= -\textbf{qq}\colon\left[\int_{\infty}d^{3}{r}\textbf{rr}\textbf{J}_{0}(\textbf{r})\right] \\
&= -\int_{\Omega}d^{3}{r}\left(\textbf{q}\cdot\textbf{r}\right)^2\textbf{J}_{0}(\textbf{r}),
\end{aligned}
\end{equation}
where we have changed the integration over all space to that of a unit cell $\Omega$ by using Eq. \ref{eqn:J0_def}. Next, we define an auxiliary tensor $\left(\textbf{q}\cdot\textbf{r}\right)^2\textbf{J}_{0}\textbf{r}$ where $\textbf{J}_{0}\textbf{r}\equiv\textbf{J}_{0}(\textbf{r})\textbf{r}$ is a dyadic tensor, which obeys
\begin{equation}
\label{eqn:aux_def}
\begin{aligned}
\nabla\cdot\left[\left(\textbf{q}\cdot\textbf{r}\right)^2\textbf{J}_{0}\textbf{r}\right] &= \nabla\left(\textbf{q}\cdot\textbf{r}\right)^2\cdot\left(\textbf{J}_{0}\textbf{r}\right) + \left(\textbf{q}\cdot\textbf{r}\right)^2 \nabla\cdot\left(\textbf{J}_{0}\textbf{r}\right)\\
&=2\left(\textbf{q}\cdot\textbf{r}\right)\left(\textbf{q}\cdot\textbf{J}_{0}\right)\textbf{r} + \left(\textbf{q}\cdot\textbf{r}\right)^2\textbf{J}_{0}.
\end{aligned}
\end{equation}
We now integrate Eq. \ref{eqn:aux_def} over an (arbitrary) volume $V$ that fully contains the unit cell $\Omega$ in Eq. \ref{eqn:quad_term}. By applying the Stokes’ theorem on the left-hand side, we obtain
\begin{equation}
\label{eqn:stokes}
\int_{V}d^{3}{r}\nabla\cdot\left[\left(\textbf{q}\cdot\textbf{r}\right)^2\textbf{J}_{0}\textbf{r}\right] = \int_{S_V}d\textbf{S}\cdot\left[\left(\textbf{q}\cdot\textbf{r}\right)^2\textbf{J}_{0}\textbf{r}\right],
\end{equation}
where $S_V$ is the boundary surface of $V$. As  there is no $\textbf{J}_{0}$ on any surface outside $\Omega$ due to Eq. \ref{eqn:J0_def}, both sides of Eq. \ref{eqn:stokes} becomes zero. Hence, an integration on the right-hand side of Eq. \ref{eqn:aux_def} over $\Omega$ is also zero, namely,
\begin{equation}
\int_{\Omega}d^{3}{r}\left[2\left(\textbf{q}\cdot\textbf{r}\right)\left(\textbf{q}\cdot\textbf{J}_{0}\right)\textbf{r} + \left(\textbf{q}\cdot\textbf{r}\right)^2\textbf{J}_{0}\right] = 0,
\end{equation}
or alternatively,
\begin{equation}
\label{eqn:stokes_result}
\begin{aligned}
\int_{\Omega}d^{3}{r}\left(\textbf{q}\cdot\textbf{r}\right)^2\textbf{J}_{0} &= \frac{2}{3} \int_{\Omega}d^{3}{r} \left[\left(\textbf{q}\cdot\textbf{r}\right)^2\textbf{J}_{0}-\left(\textbf{q}\cdot\textbf{r}\right)\left(\textbf{q}\cdot\textbf{J}_{0}\right)\textbf{r}\right] \\
&= \frac{2}{3} \int_{\Omega}d^{3}{r} \left(\textbf{q}\cdot\textbf{r}\right)\Big[\left(\textbf{r}\times\textbf{J}_{0}\right)\times\textbf{q}\Big].
\end{aligned}
\end{equation}
Combining Eq. \ref{eqn:stokes_result} with Eqs. \ref{eqn:Abar_taylor} and \ref{eqn:quad_term}, we arrive at
\begin{equation}
\label{eqn:Abar_final_J0}
\overline{\textbf{A}} = -\frac{\mu_{0}}{\Omega}  \hat{\textbf{q}}\cdot \Big \{ \frac{1}{3} \int_{\Omega}d^{3}{r} \textbf{r}\left(\textbf{r}\times\textbf{J}_{0}\right)\Big \} \times\hat{\textbf{q}},
\end{equation}
where $\hat{\textbf{q}}$ is the unit vector normal to the surface, along which the limit $\textbf{q}\to\textbf{0}$ is taken. Note that the quantity inside the curly braces in Eq. \ref{eqn:Abar_final_J0} is nothing but the magnetic quadrupole tensor\cite{Dubovik,Gray}
\begin{equation}
\label{eqn:MQUAD_def}
\overleftrightarrow{\textbf{M}} = \frac{1}{3} \int_{\Omega}d^{3}{r} \textbf{r}\left(\textbf{r}\times\textbf{J}_{0}\right)=\frac{1}{3} \int_{\Omega}d^{3}{r} \textbf{r}\textbf{m}\left(\textbf{r}\right),
\end{equation}
where $\textbf{m}\left(\textbf{r}\right)$ is the density of the magnetic moment of the bulk system. Note that, despite that the microscopic current density $\textbf{J}_{0}$ in our derivation is only phenomenological, the final expression for $\overline{\textbf{A}}$ depends only on physical observables, namely, the magnetic moments. Summarizing the derivations from Eq. (5) to Eq. (21), we finally obtain
\begin{equation}
\label{eqn:Abar_final_M}
\overline{\textbf{A}} = -\frac{\mu_{0}}{\Omega}  \hat{\textbf{q}}\cdot \overleftrightarrow{\textbf{M}} \times\hat{\textbf{q}}.
\end{equation}

\begin{figure}[tbp]
\includegraphics[width=0.95\columnwidth]{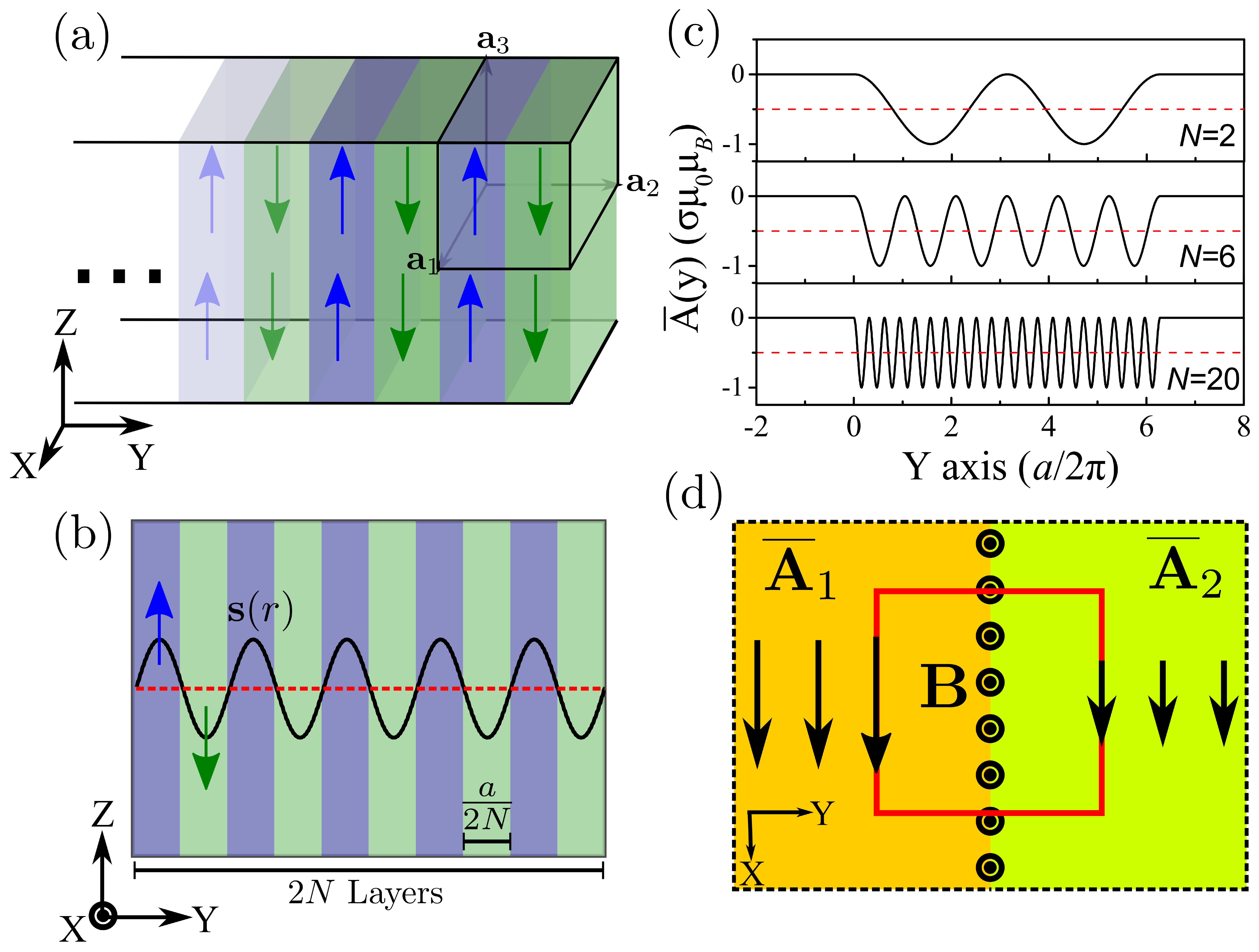}
\caption{\label{fig:fig3} (Color online) (a) A model of layered antiferromagnet with opposite spins colored in blue and green respectively. the cubic frame shows the minimum magnetic unit cell. (b) An antiferromagnetic slab with a sinusoidal spin density and a wavelength of $a/N$. The slab exhibits a thickness of $a$ and a layer number of $N$. (c) Planar-averaged vector potential $\overline{\textbf{A}}(y)$ for different $N$ calculated by a direct integration over magnetic moments induced vector potential. The average value $\overline{\textbf{A}}$ for bulk is denoted by the red dashed line. (d) A schematic illustration of the macroscopic interfacial magnetic field $\textbf{B}$. Red rectangle is the integration loop over $\overline{\textbf{A}}$ from which one obtains the interfacial magnetic flux. Symbol $\odot$ denotes the net flux which is along positive z axis.}
\end{figure}

Currently, first-principles study of magnetic systems can be challenging, thus instead we apply Eq. \ref{eqn:Abar_final_M} in two model spin systems with discrete and continuous spin distributions, respectively. Taking into account the quantum mechanical effect of spins as interpreted in Ref. \cite{Spaldin}, the spin magnetic quadruple yields
\begin{equation}
\label{eqn:QM_quad}
\overleftrightarrow{\textbf{M}} = g\mu_{B} \int_{\Omega}d^{3}{r} \textbf{r}\textbf{s}\left(\textbf{r}\right),
\end{equation}
where $g=2$ is the spin structure factor. 

The first example is the toy model shown in Fig. 3(a) which is a simple layered antiferromagnetic structure in which the magnetic unit cell contains two layers, each with opposite spins. For simplicity, we assume that the cell is cubic with three lattice vectors $|\textbf{a}_{1,2,3}|=a$, and two spin momenta $\pm \textbf{s}_0$ are localized at $\left(\frac{1}{2}a, \frac{1}{4}a, \frac{1}{2}a\right)$ and $\left(\frac{1}{2}a, \frac{3}{4}a, \frac{1}{2}a\right)$, respectively. We cut the bulk along the $xz$ plane with the vector $\hat{\textbf{q}} = \textbf{e}_{y}$ normal to the surface, leaving the moment in the outmost layer to be spin down. Upon plugging Eq. \ref{eqn:QM_quad} into Eq. \ref{eqn:Abar_final_M}, for  $\hat{\textbf{q}} = \textbf{e}_{y}$  we obtain
\begin{equation}
\overline{\textbf{A}} = -\frac{\mu_{0}}{2\Omega}g\mu_{B}{s_0}a \textbf{e}_{x}.
\end{equation}
We can recognize that the characteristic of average vector potential being discontinuous at surfaces is captured even within this minimal model.

Going beyond a collection of classical point dipoles, we introduce in the second model a continuous spin density as shown in Figs. 3(b). The spin density reads
\begin{equation}
\textbf{s}(r) = \sigma\frac{N\pi}{2a}\text{sin}\left(\frac{2N\pi}{a}y\right) \textbf{e}_{z},
\end{equation}
where $2N$ planar ferromagnetic monolayers, of thickness $a/2N$ each, are arranged in an antiferromagnetic fashion in the $y$ direction with $\sigma$ being the in-plane spin density. This configuration can be easily achieved in standard multilayer magnetic structures such as NiCl$_2$\cite{Jiang}, CrI$_3$\cite{Huang,Sivadas}, and MnBi$_2$Te$_4$\cite{LiJH,LiY,Chen}. For every slice in Figure 3(b), the corresponding contribution to planer-averaged vector potential can be calculated with Eq. \ref{eqn:A_plane}, then we sum over the whole slab to obtain the total planer-averaged vector potential $\overline{\textbf{A}}(y)$. Figure 3(c) shows the calculated $\overline{\textbf{A}}(y)$ for different layer numbers. While the planar averaged potential outside is strictly a constant, the potential inside the slab has a sinusoidal variation. It is interesting to note that no matter how one alters $N$, the maximum and minimum values for $\overline{\textbf{A}}(y)$ remain unchanged. The magnitude of $\overline{\textbf{A}}$ is, on the other hand, calculated to be $-\frac{\mu_{0}}{2}\sigma\mu_{B}$, which is exactly the same as the value obtained from Eq. \ref{eqn:Abar_final_M} using the result in Eq. \ref{eqn:QM_quad}. As $N\rightarrow\infty$ (with thickness $a$ fixed), the system should approach a macroscopic system. In such a macroscopic limit, the results in Fig. 3(c) are not expected to change.

Because $\textbf{B}=\nabla\times\textbf{A}$, the (average) macroscopic magnetic field $\overline{\textbf{B}}$ for $\textbf{A}$ may be expressed as
\begin{equation}
\overline{\textbf{B}} = \frac{\mu_{0}}{4\pi\Omega} \int_{\Omega}d^{3}{r} \nabla_{\textbf{r}}\times \int_{\infty}d^{3}{r'} \frac{\textbf{J}(\textbf{r}')}{|\textbf{r}-\textbf{r}'|}.
\end{equation}
Using the same approach in Eqs. (5-9), one may rewrite Eq. (26) as
\begin{equation}
\overline{\textbf{B}} = \frac{\mu_{0}}{4\pi\Omega} \int_{\infty}d^{3}{r} \nabla_{\textbf{r}}\times \int_{\infty}d^{3}{r'} \frac{\textbf{J}_{0}(\textbf{r}')}{|\textbf{r}-\textbf{r}'|},
\end{equation}
which allows for a Fourier transform and Taylor expansion in analogy to Eq. \ref{eqn:Abar_taylor} to arrive at
\begin{equation}
\begin{aligned}
\overline{\textbf{B}} = \frac{\mu_{0}i}{\Omega} \lim_{\textbf{q}\to\textbf{0}} \frac{\textbf{q}}{|\textbf{q}|^2} \times \Big \{ \textbf{J}_{0}(\textbf{0}) + \textbf{q}\cdot\left[\nabla\textbf{J}_{0}(\textbf{q})\right]|_{\textbf{q}=\textbf{0}} + O(q^2) \Big \}.
\end{aligned}
\end{equation}
Evidently, $\overline{\textbf{B}}$ in bulk is zero, as there are no net monopoles or dipoles in an antiferromagnetic system. At the interface, however, things will be different because $\textbf{B}=\nabla\times\textbf{A}$ is a divergence; a finite magnetic field will emerge here due to the discontinuity in $\overline{\textbf{A}}$. By choosing an integration loop straddling the interface as shown in Fig. 3(d), one will obtain a finite magnetic flux confined at the interface.

Our theory provides an alternative to surface magnetization and external magnetic field observed in magnetoelectric antiferromagnetic materials. In the conventional theory, it was argued that the symmetry requirement for equilibrium boundary state will trigger the relaxation of a magnetic configuration at the surface\cite{Belashchenko,Wu,Andreev} to result in a finite magnetization at the surface and a magnetic field outside with a 1/R$^4$ dependence. In our theory, the trigger becomes completely natural, as it is nothing but the non-vanishing bulk quadrupole. Additional relaxations are warranted as a many-body response to such an intrinsic triggering as the ultimate driving force for the effect.

Before closing, we would also like to compare the magnetic quadrupole in Eq. \ref{eqn:MQUAD_def} with the electric quadrupole defined as\cite{Choe2019}
\begin{equation}
\overleftrightarrow{\textbf{Q}} = \frac{1}{2} \int_{\Omega}d^{3}{r} \textbf{rr}\rho\left(\textbf{r}\right),
\end{equation}
where $\rho\left(\textbf{r}\right)$ is the charge density. It is known that, upon a spatial inversion operation ($\mathcal{P}$), $\textbf{r}$ changes sign but $\textbf{m}\left(\textbf{r}\right)$ does not. Conversely, upon a time-reversal operation ($\mathcal{T}$), $\textbf{r}$ does not change sign but $\textbf{m}\left(\textbf{r}\right)$ does. Hence, for a system that preserves the inversion symmetry, one has
\begin{equation}
\begin{aligned}
\overleftrightarrow{\textbf{Q}} &= \mathcal{P}\overleftrightarrow{\textbf{Q}} = \overleftrightarrow{\textbf{Q}} \\
\overleftrightarrow{\textbf{M}} &= \mathcal{P}\overleftrightarrow{\textbf{M}} = -\overleftrightarrow{\textbf{M}}.
\end{aligned}
\end{equation}
For a system that preserves the time-reversal symmetry, on the other hand, one has
\begin{equation}
\begin{aligned}
\overleftrightarrow{\textbf{Q}} &= \mathcal{T}\overleftrightarrow{\textbf{Q}} = \overleftrightarrow{\textbf{Q}} \\
\overleftrightarrow{\textbf{M}} &= \mathcal{T}\overleftrightarrow{\textbf{M}} = -\overleftrightarrow{\textbf{M}}.
\end{aligned}
\end{equation}
The fact that $\overleftrightarrow{\textbf{M}} = -\overleftrightarrow{\textbf{M}}$ in Eqs. (30-31) suggests that the magnetic quadrupole will vanish, when the system preserves either the spatial inversion or time reversal symmetry. This may be contrasted to $\overleftrightarrow{\textbf{Q}}$ for which such a condition does not exist and hence in solids $\overleftrightarrow{\textbf{Q}}$ should be omnipresent. In order to have a non-vanishing $\overleftrightarrow{\textbf{M}}$, both the $\mathcal{P}$ and $\mathcal{T}$ symmetries have to be broken. This is exactly the symmetry requirements for the onset of a magnetoelectric effect\cite{Zarzuela} in antiferromagnetically ordered bulk materials.

In summary, in spite of their fundamental importance in quantum mechanics, magnetic vector potential is considered cumbersome and rarely used in the study of magnetism. Here, we show that such a potential plays an indispensable role in defining the interfacial magnetism, at least for antiferromagnetic layered materials. In contradiction to the common belief that magnetic moments in such systems cancel each other, we find that, when a bulk crystal is truncated, the moments at the very outmost layers are singled out and uncompensated by any of its nearest neighbors, exactly the same way as what happens in an electric system where the bulk electric quadrupole is an important source of dipole potentials at the surface. We further show that such a result may offer a physical explanation to the macroscopic external $1/R^4$ magnetic field observed outside the magnetoelectric antiferromagnetic materials.

\begin{acknowledgments}
This work was supported by the U.S. DOE Grant No. DE-SC0002623. The supercomputer time sponsored by National Energy Research Scientific Center (NERSC) under DOE Contract No. DE-AC02-05CH11231 and the Center for Computational Innovations (CCI) at RPI are also acknowledged. 
\end{acknowledgments}

\end{document}